\documentclass[conference]{IEEEtran}
\IEEEoverridecommandlockouts
\usepackage{cite}
\usepackage{amsmath,amssymb,amsfonts}
\usepackage{algorithmic}
\usepackage{graphicx}
\usepackage{textcomp}
\usepackage{booktabs}
\usepackage{xcolor}
\usepackage{multirow}
\usepackage{stfloats}

\usepackage{hyperref}
\def\BibTeX{{\rm B\kern-.05em{\sc i\kern-.025em b}\kern-.08em
    T\kern-.1667em\lower.7ex\hbox{E}\kern-.125emX}}
\begin{document}

\title{MMEDIT: A Unified Framework for Multi-Type Audio Editing via Audio Language Model}

\author{
    \IEEEauthorblockN{ 
        Ye Tao $^{\dagger1,2}$, 
        Wen Wu$^{2}$,
        Chao Zhang$^{2}$,
        Mengyue Wu$^{1}$, 
        Shuai Wang$^{3}$, 
        Xuenan Xu$^{*1,2}$
        \\[2ex] %
    }
    \IEEEauthorblockA{
        \small
        \itshape %
        $^1$MoE Key Lab of Artificial Intelligence, X-LANCE Lab, Shanghai Jiao Tong University \\
        $^2$Shanghai AI Laboratory \\
        $^3$Nanjing University
    }
    \vspace{-0.5cm}
    \thanks{$^{\dagger}$Work done partially during internship at Shanghai AI Lab.}
    \thanks{$^{*}$Corresponding author.}
}

\maketitle
\begin{abstract}

Text-guided audio editing requires modifying specific acoustic events while strictly preserving non-target content.
Despite recent progress, existing approaches remain fundamentally limited: most are restricted to elementary operations (e.g., addition and removal), while real-world scenarios require more extensive tasks.
Moreover, standard architectures in this paradigm often decouple text and audio processing, limiting the model's ability to align instructions with specific acoustic contexts.  
To overcome these constraints, we propose \textbf{MMEdit}, an ALM-driven framework for unified audio editing.
We systematically extend task definitions to cover a comprehensive range of operations, including addition, replacement, removal, reordering, and attribute modification. 
Furthermore, we construct a scalable synthesis pipeline to generate large-scale paired data with fine-grained event-level annotations. 
To capture complex editing semantics, we integrate Qwen2-Audio encoder with an MMDiT-based generator, achieving precise cross-modal alignment and localized editing.
Experiments demonstrate that our approach delivers superior editing localization accuracy, robust instruction following, and high fidelity in non-edited regions. 
The project page can be accessed at \href{https://ty0402.github.io/MMEditing/}{\textcolor{blue!60!black}{Project Page}}

\end{abstract}

\begin{IEEEkeywords}
Audio editing, Diffusion models, Multimodal models
\end{IEEEkeywords}

\section{Introduction}
\label{sec:intro}

With the rapid development of generative modeling, modern text-to-audio (TTA) systems~\cite{yang2023diffsound,liu2023audioldm,majumder2024tango} are now capable of producing high-fidelity, diverse, and text-aligned sounds from natural language prompts.
Unlike TTA, which synthesizes audio from scratch, text-guided audio editing (TTE) seeks to modify specific regions of an existing audio signal according to user-provided textual instructions, while preserving the remaining content unchanged~\cite{wang2023audit}.
This paradigm offers users a more intuitive and efficient way to achieve fine-grained control over audio content through natural language, rather than relying on manual waveform editing or expert post-processing.
Since the introduction of this task in AUDIT~\cite{wang2023audit}, audio editing has attracted increasing research interest~\cite{xu2024prompt,manor2024zero,jia2025audioeditor,paissan2024audio}. 


Existing studies on TTE can be broadly categorized into \textbf{training-free} and \textbf{training-based} paradigms.
\textbf{Training-free approaches} typically adapt pre-trained diffusion-based TTA models by inverting the denoising process to reconstruct latent noise and re-generate audio under new textual conditions.
PPAE~\cite{xu2024prompt} and AudioEditor~\cite{jia2025audioeditor} are representative methods.
Although free from additional training, these approaches rely heavily on diffusion inversion, resulting in high inference cost and a strong dependence on full captions or handcrafted masks, restricting their applicability to complex editing tasks.
Crucially, the mechanism of inversion and subsequent regeneration may compromise the fidelity to the original structure and fine details.

In contrast, \textbf{training-based frameworks} explicitly learn instruction-conditioned generation models from paired data. 
AUDIT~\cite{wang2023audit} pioneered this paradigm by training a latent diffusion model. 
Paissan \textit{et al.}~\cite{paissan2024audio} explored prompt interpolation and fine-tuning within the embedding space, while RFM-Editing~\cite{gao2025rfm} incorporated rectified flow matching to improve efficiency.  Compared to training-free inversion, this paradigm offers better fidelity and faster inference speeds.

Despite these advances, the progress of audio editing faces challenges in both \textbf{data resources} and \textbf{model design}.
In terms of \textit{data}, real editing pairs are rarely available, hence existing studies rely on small-scale or manually synthesized datasets to simulate editing operations\cite{wang2023audit,jia2025audioeditor}. 
However, prior synthetic pipelines often adopt simple mixing strategies with limited event-level controllability, resulting in coverage of only basic edit types (i.e., addition and removal), and therefore failing to reflect a comprehensive range of real-world manipulations.
The absence of an open and unified benchmark further exacerbates the challenge, leading to inconsistent evaluation and limited reproducibility across different studies. 
In terms of \textit{model design}, most existing works rely on text-only encoders to process instructions.
The interaction between textual instructions and the source audio is limited to the cross-attention mechanism in the generation model.
Yet in complex audio scenes, accurately editing a specific event entails not only interpreting the instruction but also locating the event within the source audio, a capability that text-only semantic encoders inherently lack.
Consequently, such a decoupled, late-fusion design is often inadequate for achieving context-aware understanding of complex editing instructions.



To address these limitations, we propose \textbf{MMEdit}, a unified framework driven by an Audio Language Model for general, instruction-following audio editing.
First, we systematically expand the editing scope to six representative operations: \textit{addition}, \textit{removal}, \textit{replacement}, \textit{loudness adjustment}, \textit{reordering}, and \textit{speed modification} for better adaptation with real-world applications. We thereby develop a scalable pipeline yielding over one million training samples, distilling a high-quality, human-verified benchmark for evaluation.
Although some operations can be achieved by signal processing tools in single-event audio, event overlap in realistic soundscapes makes event-specific editing difficult.
Therefore we integrate them into a unified editing model to learn shared audio editing capabilities across operations.
Moreover, MMEdit utilizes an Audio Language Model (ALM) to process instructions and audio content jointly.
This unified processing enables precise cross-modal alignment, allowing the model to precisely interpret instructions within the specific acoustic context.
The ALM encoder is integrated with an \textit{MMDiT-based} generator~\cite{esser2024scaling} to generate high-fidelity audio.
\noindent
\textbf{Our main contributions are summarized as follows:}
\begin{itemize}
    \item \textbf{Comprehensive Task Definition:} We systematically expand the editing landscape to encompass six representative operations, better reflecting real-world requirements.
    \item \textbf{Scalable Data Pipeline and Unified Benchmark:} We develop a synthetic generation pipeline that produces large-scale paired data with fine-grained, event-level annotations. To foster reproducibility, we establish and open-source an evaluation benchmark on TTE for the first time. 
    \item \textbf{ALM-Driven Architecture and Superior Performance:} MMEdit involves the ALM encoder with MMDiT to enhance the model's ability to understand complex editing instructions for precise editing. Extensive experiments demonstrate that MMEdit significantly outperforms existing baselines in following editing instructions and maintaining the integrity of unedited content.
\end{itemize}

\section{Data pipeline}
\label{sec:datapipline}
\begin{figure*}[t!]
  \centering

  \includegraphics[width=\textwidth]{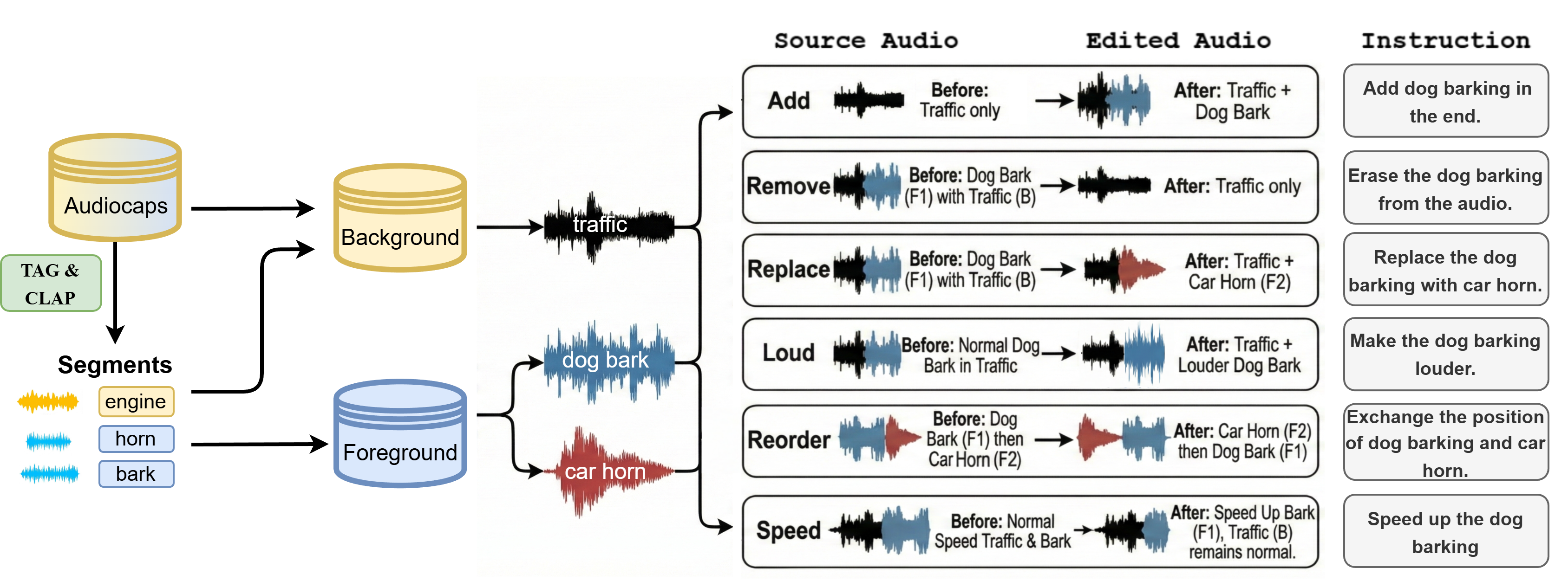}
  \caption{
  \textbf{Construction pipeline and examples.}
  (1) Extract clean event-level segments from AudioCaps using TAG model and CLAP filtering; 
  (2) Simulate foreground–background compositions and record metadata; 
  (3) Generate template-based instructions for each editing task. 
  }
  \label{fig:pipeline}
\end{figure*}

To facilitate scalable training and evaluation, 
we construct a large-scale synthetic dataset comprising over one million audio editing pairs. 
Each pair is represented as a triplet $\langle \text{instruction}, \text{audio}_{\text{source}}, \text{audio}_{\text{edited}} \rangle$.
The construction pipeline, illustrated in Fig.~\ref{fig:pipeline}, comprises three stages: segment curation, audio composition, and instruction synthesis.

\subsection{Audio Segment Curation}

Existing TTE studies like AUDIT typically synthesize training pairs by combining entire audio clips with their full global captions. 
However, to enable more precise and interpretable editing such as specifying event timing and occurrence counts, a finer granularity is required. We first perform event-level segmentation and grounding on an audio-caption corpus, AudioCaps~\cite{kim2019audiocaps}.
Specifically, inspired by AudioTime~\cite{xie2025audiotime}, we employ an LLM to decompose each caption into single-event descriptions.
The descriptions are then timestamped by the Text-to-Audio Grounding (TAG) model~\cite{xu2024towards} to determine event boundaries, and only the non-overlapping segments are retained and cropped as base audio units for subsequent synthesis.
Finally, to guarantee semantic alignment, we compute CLAP scores~\cite{elizalde2023clap} between the textual label and the extracted segment and discard pairs with a similarity score below $0.3$. 

To simulate realistic scenes with rich contextual variations, we categorize audio components into \textbf{foreground (F)} and \textbf{background (B)} sounds. 
Foreground segments typically consist of short-duration acoustic events derived from our curated segments and the AudioTime dataset. Conversely, background segments are sourced from full-length AudioCaps recordings and long-duration intervals, capturing the sustained environmental context.
This differentiation provides explicit control over the spatial and semantic hierarchy of events, forming the foundation for the following editing task definitions.


\subsection{Editing Task Definition and Composition}
\label{ssec:editing}
Building upon the disentangled foreground and background components, we systematically define six representative editing operations that simulate real-world acoustic manipulations. 
We formulate each operation as a transformation function applied to the foreground and background streams, parameterized by key attributes: \textit{event type, quantity, and temporal position}. 
In our notation, $F_n$ denotes the $n$-th foreground event, while $F'$ and $B'$ represent new foreground and background components, respectively.
To technically implement these compositional processes, we leverage the Scaper toolkit~\cite{salamon2017scaper} for programmable soundscape synthesis and augmentation.
This allows for precise control over signal-to-noise ratios (SNR), event timing, and dynamic mixing.
This unified formulation accommodates multiple subtypes within each category, enabling a flexible and scalable framework for diverse audio editing scenarios.
\subsubsection{Addition}
This task introduces new acoustic elements into an existing audio clip. 
It covers three typical settings: $(B \rightarrow F + B), (F \rightarrow F + B)$, and \((F + B \rightarrow F + F' + B)\).
These variants capture common compositional patterns in real-world audio.

\subsubsection{Removal}
This task removes specific sound components while preserving the rest of the audio. 
It includes removing a localized foreground or background event: \((F + B \rightarrow B)\),
\((F + B \rightarrow F)\), \((F + F'+ B \rightarrow F + B)\).

\subsubsection{Replacement}
This task substitutes a target sound event with another while maintaining its temporal occurrence in the original audio clip: \((F + B \rightarrow F' + B)\) or \((F + B \rightarrow F + B')\).

\subsubsection{Reordering}
This task changes the temporal sequence of multiple events without altering their content: \((F_1 + F_2  \rightarrow F_2 + F_1 )\) and \((F_1 + F_2 + B \rightarrow  F_2 + F_1 + B)\).

\subsubsection{Loudness Adjustment}
This task modulates the relative amplitude of selected event sources to simulate emphasis or attenuation while keeping all other attributes unchanged: \((F + B \rightarrow F^{\uparrow / \downarrow} + B)\).

\subsubsection{Speed Modification}
This task adjusts the playback rate of a sound event while keeping all other attributes unchanged: \((F + B \rightarrow F \times \alpha + B)\), where the scaling factor \(\alpha\) controls the speed.

%

Given the structured nature of audio editing instructions, we designed diverse prompt templates for each editing type, incorporating specific slots for variable parameters such as event types, timestamps, and occurrence counts. We then construct corresponding textual instructions by populating these slots with fine-grained metadata to simulate realistic user commands.
By scaling this pipeline, we construct a massive training corpus comprising over one million triplets, amounting to approximately 2500 hours of audio.
To ensure a reliable evaluation standard, we further curate a high-quality benchmark through rigorous screening and human verification. We make both this benchmark and our generation pipeline publicly available to foster reproducible research and facilitate fair comparison.
The detailed statistics for each editing task are presented in Table~\ref{tab:dataset_stats}.

\begin{table}[t!]
    \centering
    \caption{\textbf{Statistics of MMEdit Data.} The distribution of training samples and the benchmark across editing tasks.}
    \label{tab:dataset_stats}
    \setlength{\tabcolsep}{3.5pt} 
    \begin{tabular}{lcccccc}
        \toprule
        \textbf{Split} & \textbf{Add} & \textbf{Remove} & \textbf{Replace} & \textbf{Reorder} & \textbf{Loud.} & \textbf{Speed} \\
        \midrule
        Train & 260k & 230k & 250k & 150k & 100k & 100k \\
        Benchmark  & 700 &  500 & 300  & 500  & 300 & 200 \\
        \bottomrule
    \end{tabular}
\end{table}

\section{MMEDIT}
\label{sec:method}

In this section, we introduce \textbf{MMEdit}, a general-purpose framework designed for instruction-following audio editing.
Distinct from prior decoupled designs that rely on cross-attention to bridge modalities, MMEdit integrates an ALM to process instructions and audio jointly for deep alignment, combined with an MMDiT backbone~\cite{esser2024scaling} for high-fidelity generation.
This design enables more precise cross-modal alignment, allowing the model to interpret instructions within the specific acoustic context.
As illustrated in Fig.~\ref{fig:architecture}, MMEdit consists of three major components:
a waveform variational autoencoder (VAE)~\cite{xu2025uniflow} for latent audio representation,
an Audio Language Model (ALM)-based encoder for cross-modal understanding,
and an MMDiT-based backbone that performs generation in the latent space.

\begin{figure*}[t]
    \centering
    \includegraphics[width=\textwidth]{fig/mmedit_method.png}
    \caption{
    \textbf{Overall architecture of the proposed MMEdit framework.}
    The model employs the powerful Qwen2-Audio for joint understanding of audio and instruction, generating the high-level multi-modal representation $\mathbf{H}$.
    Then the MMDiT backbone predicts the audio latent $\mathbf{z}_0$ from a Gaussian latent conditioned on $\mathbf{H}$ and the latent of the source audio $\mathbf{z}_\mathrm{in}$.
    Finally, the VAE decoder converts $\mathbf{z}_\mathrm{0}$ to the edited audio.
    }
    \label{fig:architecture}
\end{figure*}

\subsection{Multimodal Understanding}
\label{subsec:audiollm}
Crucial to general editing capabilities is the model's ability to deeply understand the interaction between the source audio content and the textual instruction. 
Instead of relying on a text-only encoder (e.g., T5) for instruction understanding, we use \textit{Qwen2-Audio}~\cite{chu2024qwen2} as a multimodal encoder to obtain a more representative semantic feature to guide editing. 
Qwen2-Audio can jointly process the source audio $\mathbf{x}_{in}$ and instruction $\mathbf{y}$, yielding inherently cross-modal aligned representations:
\begin{equation}
    \mathbf{H} = \mathcal{E}_{\text{Qwen}}(\mathbf{x}_{in}, \mathbf{y})
    \in \mathbb{R}^{L_q \times D_q}.
\end{equation}
From $\mathbf{H}$, we derive two complementary conditioning signals:
\begin{itemize}
    \item \textbf{Global Context:}
    \(
    \mathbf{c}_{\text{global}} = \text{AvgPool}(\mathbf{H})
    \),
    summarizing the holistic intent of the editing.
    \item \textbf{Fine-grained Features:}
    \(
    \mathbf{H}_{\text{seq}} = \mathbf{H}
    \),
    providing event-level temporal cues and enabling the backbone to perform precise localization through joint attention.
\end{itemize}

This dual-granularity conditioning, motivated by MMDiT, benefits from offering both global summaries and fine-grained local details as complementary conditioning signals.

\subsection{The MMDiT Backbone Architecture}
To effectively process the composite input under complex multimodal guidance, we build upon the MMDiT architecture proposed in Stable Diffusion 3, and adapt it to audio editing by allowing audio latents to interact with multimodal instruction embeddings $\mathbf{H}$.

\subsubsection{Global Conditioning via Adaptive Modulation}
The global context $\mathbf{c}_{global}$ from the ALM is fused with the diffusion timestep $\mathbf{t}$.
The resulting vector modulates the entire network through adaptive Layer Normalization (AdaLN)~\cite{peebles2023scalable} before each Transformer block, serving as a global semantic guidance for generation.

\subsubsection{MMDiT Blocks Design}

The backbone comprises a stack of $N_{\text{joint}}$ \textbf{Joint Blocks} followed by $N_{\text{single}}$ \textbf{Single Blocks}.
In each joint block, the audio latents and the multimodal instruction embeddings are concatenated along the sequence dimension and processed by a shared self-attention layer.
This joint attention mechanism enables latent audio directly attend to instruction embeddings, enabling event-level, text-guided editing.
After joint blocks, single blocks operate purely within audio latents using self-attention. 
Single blocks focus on refining temporal continuity and spectral coherence, ensuring that the edited audio remains acoustically natural and that unedited regions are preserved with high fidelity.

\subsection{Latent Diffusion for Audio Editing}
Given an editing pair, we denote the source waveform as
$\mathbf{x}_{\mathrm{in}}$ and the target waveform as
$\mathbf{x}_{0}$.
Both are then mapped into the latent space of a pre-trained VAE operating directly on raw waveforms:
\begin{equation}
    \mathbf{z}_{\mathrm{in}} = \mathcal{E}_{\mathrm{VAE}}(\mathbf{x}_{\mathrm{in}}),
    \qquad
    \mathbf{z}_0 = \mathcal{E}_{\mathrm{VAE}}(\mathbf{x}_{0})
    \in \mathbb{R}^{C \times L},
\end{equation}
where $\mathcal{E}_{\mathrm{VAE}}$ is the encoder of the pre-trained VAE.

During training, we follow the standard DDPM~\cite{ho2020denoising} formulation.
For a randomly sampled timestep $t \in \{1,\dots,T\}$, a noisy version $\mathbf{z}_t$ of the target latent is obtained by
\begin{equation}
    \mathbf{z}_t
    =
    \sqrt{\bar{\alpha}_t}\,\mathbf{z}_0
    +
    \sqrt{1-\bar{\alpha}_t}\,\boldsymbol{\varepsilon},
    \qquad
    \boldsymbol{\varepsilon}\sim\mathcal{N}(\mathbf{0}, \mathbf{I}),
\end{equation}
where $\bar{\alpha}_t$ is the cumulative product of the forward-diffusion coefficients $\alpha_t$, which form the noise schedule, specifying how much signal is preserved at each diffusion step.

A key difficulty in editing is to modify only the requested events while preserving the remaining structure of the input audio.
To expose the unedited reference to the denoiser, we
concatenate the noisy latent with the source audio latent along the channel dimension,
\begin{equation}
    \tilde{\mathbf{z}}_t
    =
    \mathbf{z}_t \oplus \mathbf{z}_{\mathrm{in}}
    \in \mathbb{R}^{2C \times L},
\end{equation}
where $\oplus$ denotes channel-wise concatenation.
The MMDiT backbone then predicts the velocity as
\begin{equation}
    \hat{\mathbf{v}}
    =
    \mathbf{v}_\theta\big(
        \tilde{\mathbf{z}}_t,\,
        t,\,
        \mathbf{c}_{\mathrm{global}},\,
        \mathbf{H}_{\mathrm{seq}}
    \big),
\end{equation}
with
\begin{equation}
    \mathbf{v}
    =
    \sqrt{\bar{\alpha}_t}\,\boldsymbol{\varepsilon}
    -
    \sqrt{1-\bar{\alpha}_t}\,\mathbf{z}_0,
    \qquad
    \boldsymbol{\varepsilon}\sim\mathcal{N}(\mathbf{0},\mathbf{I}).
\end{equation}
We train the model with a velocity-prediction objective:
\begin{equation}
    \mathcal{L}_{\text{diff}}
    =
    \mathbb{E}_{\mathbf{z}_0,\,\boldsymbol{\varepsilon},\,t,\,\mathbf{z}_{\mathrm{in}}}
    \Big[
        \big\|
            \mathbf{v}
            -
            \mathbf{v}_\theta(
                \tilde{\mathbf{z}}_t,\,
                t,\,
                \mathbf{c}_{\mathrm{global}},\,
                \mathbf{H}_{\mathrm{seq}}
            )
        \big\|_2^2
    \Big].
\end{equation}

During training, text instructions are randomly masked to learn unconditional generation while source audio latent is always fed to the model.
During inference, we use DDIM~\cite{songdenoising} sampler with classifier-free guidance, where the guided noise estimate is formulated as $\tilde{\epsilon} = \epsilon_{\text{uncond}} + w (\epsilon_{\text{cond}} - \epsilon_{\text{uncond}}).$
Finally, the VAE decoder converts the denoised latent into the edited audio.

\section{Experiments}

\subsection{Training Setup}
\label{subsec:exp_setup}

MMEdit is trained on the synthetic editing pairs introduced in Sec.~\ref{sec:datapipline}, together with additional text-to-audio style pairs constructed from AudioCaps. 
In preliminary experiments, we observe that mixing in TTA data
(generation from captions without edits) noticeably improves the perceptual quality of edited outputs.
Throughout training, the Qwen2-Audio encoder and the VAE are kept frozen, while the
MMDiT backbone is trained  for 80 epochs using the AdamW
optimizer.

\subsection{Baselines \& Evaluation.}

We compare MMEdit with two representative audio editing systems:
\textit{AUDIT} and \textit{AudioEditor}, since they are the only publicly available models similar to our setting.
Note that both only support Addition, Removal, and Replacement operations.
Since AudioEditor operates on source–target caption pairs, we rewrite each editing instruction into a full-sentence target caption to align with its required input format.
To comprehensively assess performance, we report both reference-based objective metrics on synthetic pairs and Mean Opinion Score (MOS)-based subjective metrics on real recordings.

\subsubsection{\textbf{Objective evaluation}} On the synthetic test set,
we report Log-Spectral Distance (LSD), Fr\'echet Audio Distance (FAD), Fr\'echet Distance (FD), and KL divergence to measure spectral and distributional discrepancies, alongside Inception Score (IS)~\cite{liu2023audioldm} for perceptual quality and diversity. 
\subsubsection{\textbf{Subjective evaluation}}
\label{subsec:subjective}
To mitigate the potential bias inherent in synthetic data, we conduct a listening test on a curated subset of 60 real \textit{AudioCaps} recordings paired with manually designed instructions.
As no ground-truth edited audio is available, we rely on
human judgment.
For all tasks, we invite ten college-level raters to do a subjective rating with explicit instructions.
Each edited sample is rated on a 1–5 Likert scale along two dimensions:
\emph{Relevance} (R-MOS), which measures how accurately the system follows the
editing instruction;
and \emph{Faithfulness} (F-MOS), which measures how well non-edited regions of the
original audio are preserved without introducing unintended artifacts or distortions.

\begin{table*}[t]
\centering
\setlength{\tabcolsep}{5.2pt}
\renewcommand{\arraystretch}{1.12}
\caption{\textbf{Quantitative comparisons. }
We report objective metrics (LSD, FAD, FD, KL, IS) on the synthetic test set and subjective MOS ratings on real-world recordings.
$\downarrow$ indicates lower is better, while $\uparrow$ indicates higher is better.
Note that \textit{AudioEditor} and \textit{AUDIT} do not support Reordering, Loudness, or Speed operations.
}
\label{tab:editing-main}
\begin{tabular}{l l | c c c c c | c c}
\toprule
\textbf{Editing Type} & \textbf{Model} &
\textbf{LSD} $\downarrow$ &
\textbf{FAD} $\downarrow$ &
\textbf{FD} $\downarrow$ &
\textbf{KL} $\downarrow$ &
\textbf{IS} $\uparrow$ &
\textbf{R-MOS} $\uparrow$ &
\textbf{F-MOS} $\uparrow$
\\
\midrule
\multirow{3}{*}{Add}
  & AUDIT         & 1.962 & 2.652 & 24.573 & 2.073 & 5.050 & 2.79&2.67 \\ 
  & AudioEditor          & 1.982 & 3.125 & 23.351 & 1.993 & \textbf{10.492} & 2.79 & 3.25\\
  & \textbf{MMEdit} & \textbf{1.614} & \textbf{0.826} & \textbf{15.536} & \textbf{1.247} & 8.348 & \textbf{4.34} & \textbf{4.09} \\
\midrule
\multirow{3}{*}{Remove}
  & AUDIT        & 3.432 & 4.138 & 30.461 & 2.456 & 4.675 & 2.84 & 1.82 \\
  & AudioEditor          & 3.780 & 5.747 & 31.370 & 2.013 & \textbf{10.025} &2.70&2.70 \\
  & \textbf{MMEdit} & \textbf{2.896} & \textbf{0.678} & \textbf{13.080} & \textbf{0.690} & 9.187 & \textbf{3.90} & \textbf{4.30} \\
\midrule
\multirow{3}{*}{Replace}
  & AUDIT       & 2.850 & 3.145 & 28.126 & 2.209 & 3.600 & 2.84 & 3.75 \\
  & AudioEditor          & 2.629 & 2.834 & 31.994 & 1.764 & \textbf{7.905} & 2.70&2.70 \\
  & \textbf{MMEdit} & \textbf{2.560} & \textbf{0.946} & \textbf{19.060} & \textbf{1.760} & 6.080  & \textbf{3.73} & \textbf{4.19} \\
\midrule
\multirow{1}{*}{Reorder}
  & \textbf{MMEdit} & 2.353 & 0.791 & 9.328 & 0.391 & 6.381 & 3.90 & 4.40\\
\midrule
\multirow{1}{*}{Loudness}
  & \textbf{MMEdit} & 1.310 & 0.447 & 8.528 & 0.252 & 6.321 & 4.43 &4.35\\
\midrule
\multirow{1}{*}{Speed}
  & \textbf{MMEdit}& 1.209 & 0.479 &8.632& 0.273&5.575 & 3.88 & 4.18 \\
\bottomrule
\end{tabular}
\end{table*}

\subsection{Main Results and Analysis}
\label{subsec:main_results_text}

Table~\ref{tab:editing-main} summarizes the quantitative results on the
synthetic test set for all tasks.
Across three tasks that all models support, MMEdit consistently achieves the best result except for IS, indicating that the edited outputs are closer to the ground-truth
targets. 
In contrast, AUDIT shows noticeably higher distance and divergence values, suggesting degraded sample fidelity and weaker alignment with the ground-truth.
AudioEditor achieves the highest IS, while MMEdit attains slightly lower yet competitive values and AUDIT trails behind both.
This outcome is anticipated, given that AudioEditor operates as an inversion-based framework built upon a TTA generation backbone.
Since each editing operation necessitates triggering the full generation process of the backbone, it inherently promotes greater diversity in resynthesis, yielding a higher IS.
Conversely, this reliance on holistic regeneration accounts for its inferior performance on similarity-oriented metrics, indicating a reduced capacity to preserve the unedited content of the source audio compared to MMEdit.


As detailed in Table~\ref{tab:editing-main}, subjective results on real AudioCaps clips also confirm MMEdit's superiority.
MMEdit consistently scores above 3.7 in R-MOS, indicating strong instruction following.
This advantage is even more pronounced in F-MOS, where MMEdit demonstrates superior preservation of non-edited regions. The result further validated by paired $t$-tests ($p < 0.001$) against both baselines.
Notably, despite high IS, AudioEditor receives markedly lower subjective scores, suggesting more aggressive modifications to maximize
diversity at the cost of destroying non-editing
contents. 


Furthermore, MMEdit demonstrates robust generalization by executing complex operations such as Reordering, Loudness, and Speed adjustments—tasks that existing baselines fail to support.
These results conclusively highlight the superiority of our framework: MMEdit achieves a significantly more favorable balance between distributional faithfulness and generation diversity compared to prior works.

\subsection{Ablation Studies}

We investigate individual contributions of the ALM encoder, the diffusion backbone architecture, and the classifier-free guidance strategy~\cite{ho2021classifier}. 

\subsubsection{Effect of ALM Encoder}
To assess the critical role of the multimodal ALM, we conduct an ablation study by replacing the Qwen2-Audio encoder with a  text-only encoder (Flan-T5~\cite{chung2024scaling}), keeping other components invariant. 
As presented in Table~\ref{tab:ablation}, although the text-only variant (\textit{w/ T5}) achieves higher IS, the decline in fidelity metrics indicates that the text-only encoder fails to capture the intricate acoustic cues of the source audio.
In contrast, Qwen2-Audio contributes richer cross-modal information by jointly encoding the instruction and the source audio, enabling more accurate and perceptually coherent edits than a text-only encoder could provide.

\begin{table}[t]
    \centering
    \small
    \setlength{\tabcolsep}{4pt} 
    \caption{\textbf{Ablation study on the instruction encoder and generation backbone.}}
    \label{tab:ablation}
    \begin{tabular}{l ccc cc}
        \toprule
        \multirow{2.5}{*}{\textbf{Method}} & \multicolumn{3}{c}{\textbf{Objective}} & \multicolumn{2}{c}{\textbf{Subjective}} \\
        \cmidrule(lr){2-4} \cmidrule(lr){5-6} 
         & FD $\downarrow$ & KL $\downarrow$ & IS $\uparrow$ & R-MOS $\uparrow$ & F-MOS $\uparrow$ \\
        \midrule
        w/ T5 & 6.29 & 1.25 & \textbf{9.10} & 3.36 & 3.63 \\
        w/ DiT & 4.77 & 1.08 & 8.72 & 3.80 & 4.05\\
        \midrule
        \textbf{MMEdit (Full)} & \textbf{4.72} & \textbf{0.83} & 8.83 & \textbf{4.05} & \textbf{4.23} \\
        \bottomrule
    \end{tabular}
    \vspace{-10pt} %
\end{table}

\subsubsection{Effect of Backbone Architecture}
We further evaluate the advantage of our MMDiT backbone against the vanilla DiT variant (denoted as \textit{w/ DiT}).
This variant employs a standard DiT structure that injects $\mathbf{H}$ solely via cross-attention layers.
As shown in Table~\ref{tab:ablation}, this variant falls short of the full MMEdit framework across all metrics.
This result suggests that shallow cross-attention–based conditioning may not fully exploit the multimodal conditions.
Conversely, the MMDiT backbone offers a more effective way to integrate these cues with its joint-attention mechanism.

\subsubsection{Analysis of Guidance Scale and Condition Masking}

\begin{figure}[t]
    \centering
    \begin{minipage}{0.49\linewidth}
        \centering
        \includegraphics[width=\linewidth]{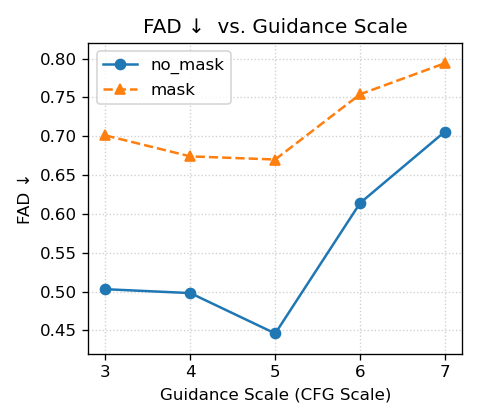} 
        \label{fig:fd_scale}
    \end{minipage}
    \hfill 
    \begin{minipage}{0.49\linewidth}
        \centering
        \includegraphics[width=\linewidth]{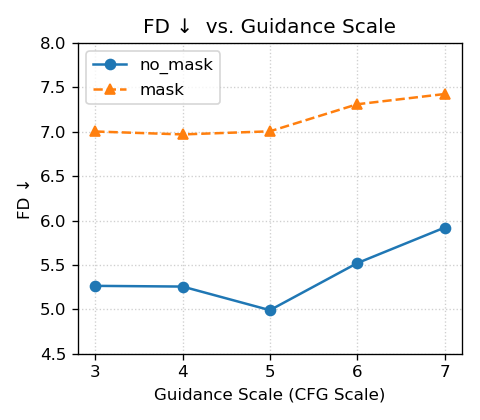} %
        \label{fig:fad_scale}
    \end{minipage}
    \caption{Effect of guidance scale and condition masking.}
    \label{fig:cfg_analysis}
    \vspace{-10pt}
\end{figure}

We finally investigate the impact of the Classifier-Free Guidance (CFG) scale $w$ and the masking strategy to train unconditional generation. 
A critical design choice lies in the strategy to train the unconditional generation.
We explore two strategies: 
(1) \textit{Masking Audio}: Dropping both the text instruction and the source audio latent; 
(2) \textit{No-Masking Audio}: Dropping only the text instruction while retaining the source audio latent.

As illustrated in Figure~\ref{fig:cfg_analysis}, \textit{No-Masking} consistently outperforms \textit{Masking} by a large margin.
The theoretical justification is that, unlike pure generation, the editing task requires preserving the underlying structure of the source audio. 
By retaining the source latent in the unconditional branch, the guidance direction is driven to focus on the semantic modifications specified by the text, rather than reconstructing the audio from scratch. 
The results indicate that excessively large scales overly distort the source audio, whereas small scales lead to insufficient editing.
A scale of $w=5.0$ achieves the most favorable trade-off between preservation and editing strength.

\section{Conclusion}
\label{sec:conclusion}

In this paper, we propose MMEdit, an ALM-driven framework designed for general-purpose text-guided audio editing. 
By constructing a scalable synthetic pipeline with fine-grained event-level annotations, we systematically extend the editing landscape to cover six distinct tasks. 
Our proposed architecture synergizes the Qwen2-Audio multimodal encoder with an MMDiT-based backbone, effectively bridging the gap between high-level semantic understanding and precise acoustic manipulation. 
Extensive experiments demonstrate that MMEdit significantly outperforms existing baselines, achieving superior instruction following and editing localization while preserving the high fidelity of non-edited regions. 
However, limitations remain regarding generation diversity and fine-grained temporal alignment, particularly for complex overlapping events, which we leave for future work.
We hope this work serves as a solid and reproducible foundation for future research into audio editing systems.

\bibliographystyle{IEEEbib}
\bibliography{icme2026references}

@string{icassp="IEEE International Conference on Acoustics, Speech, and Signal Processing"}

@string{interspeech="Annual Conference of the International Speech Communication Association"}

@string{taslp="IEEE/ACM Transactions on Audio, Speech, and Language Processing"}

@string{nips="Advances in Neural Information Processing Systems"}

@string{icml="International Conference on Machine Learning"}

@article{wang2023audit,
  title={Audit: Audio editing by following instructions with latent diffusion models},
  author={Wang, Yuancheng and Ju, Zeqian and Tan, Xu and He, Lei and Wu, Zhizheng and Bian, Jiang and others},
  journal={nips},
  volume={36},
  pages={71340--71357},
  year={2023}
}

@inproceedings{esser2024scaling,
  title={Scaling Rectified Flow Transformers for High-Resolution Image Synthesis},
  author={Esser, Patrick and Kulal, Sumith and Blattmann, Andreas and Entezari, Rahim and M{\"u}ller, Jonas and Saini, Harry and Levi, Yam and Lorenz, Dominik and Sauer, Axel and Boesel, Frederic and others},
  booktitle={icml},
  year={2024}
}

@article{xu2024towards,
  title={Towards weakly supervised text-to-audio grounding},
  author={Xu, Xuenan and Ma, Ziyang and Wu, Mengyue and Yu, Kai},
  journal={IEEE Transactions on Multimedia},
  year={2024},
  publisher={IEEE}
}

@article{yang2023diffsound,
  title={Diffsound: Discrete diffusion model for text-to-sound generation},
  author={Yang, Dongchao and Yu, Jianwei and Wang, Helin and Wang, Wen and Weng, Chao and Zou, Yuexian and Yu, Dong},
  journal=taslp,
  volume={31},
  pages={1720--1733},
  year={2023},
  publisher={IEEE}
}

@inproceedings{majumder2024tango,
  title={Tango 2: Aligning Diffusion-based Text-to-Audio Generations through Direct Preference Optimization},
  author={Majumder, Navonil and Hung, Chia-Yu and Ghosal, Deepanway and Hsu, Wei-Ning and Mihalcea, Rada and Poria, Soujanya},
  pages={564--572},
  year={2024},
  booktitle=mm
}

@article{chung2024scaling,
  title={Scaling instruction-finetuned language models},
  author={Chung, Hyung Won and Hou, Le and Longpre, Shayne and Zoph, Barret and Tay, Yi and Fedus, William and Li, Yunxuan and Wang, Xuezhi and Dehghani, Mostafa and Brahma, Siddhartha and others},
  journal={Journal of Machine Learning Research},
  volume={25},
  number={70},
  pages={1--53},
  year={2024}
}

@inproceedings{xu2024prompt,
  title={Prompt-guided Precise Audio Editing with Diffusion Models},
  author={Xu, Manjie and Li, Chenxing and Zhang, Duzhen and Su, Dan and Liang, Wei and Yu, Dong},
  booktitle={icml},
  pages={55126--55143},
  year={2024},
  organization={PMLR}
}

@inproceedings{jia2025audioeditor,
  title={Audioeditor: A training-free diffusion-based audio editing framework},
  author={Jia, Yuhang and Chen, Yang and Zhao, Jinghua and Zhao, Shiwan and Zeng, Wenjia and Chen, Yong and Qin, Yong},
  booktitle={icassp},
  pages={1--5},
  year={2025},
  organization={IEEE}
}

@incollection{paissan2024audio,
  title={Audio Editing with Non-Rigid Text Prompts},
  author={Paissan, Francesco and Della Libera, Luca and Wang, Zhepei and Ravanelli, Mirco and Smaragdis, Paris and Subakan, Cem and others},
  booktitle={interspeech},
  year={2024}
}

@inproceedings{manor2024zero,
  title={Zero-shot unsupervised and text-based audio editing using DDPM inversion},
  author={Manor, Hila and Michaeli, Tomer},
  booktitle={Proceedings of the 41st International Conference on Machine Learning},
  pages={34603--34629},
  year={2024}
}

@article{chu2024qwen2,
  title={Qwen2-audio technical report},
  author={Chu, Yunfei and Xu, Jin and Yang, Qian and Wei, Haojie and Wei, Xipin and Guo, Zhifang and Leng, Yichong and Lv, Yuanjun and He, Jinzheng and Lin, Junyang and others},
  journal={arXiv preprint arXiv:2407.10759},
  year={2024}
}

@article{gao2025rfm,
  title={RFM-Editing: Rectified Flow Matching for Text-guided Audio Editing},
  author={Gao, Liting and Yuan, Yi and Chen, Yaru and Cheng, Yuelan and Li, Zhenbo and Wen, Juan and Zhang, Shubin and Wang, Wenwu},
  journal={arXiv preprint arXiv:2509.14003},
  year={2025}
}

@inproceedings{xie2025audiotime,
  title={Audiotime: A temporally-aligned audio-text benchmark dataset},
  author={Xie, Zeyu and Xu, Xuenan and Wu, Zhizheng and Wu, Mengyue},
  booktitle={icassp},
  pages={1--5},
  year={2025},
  organization={IEEE}
}

@inproceedings{kim2019audiocaps,
  title={Audiocaps: Generating captions for audios in the wild},
  author={Kim, Chris Dongjoo and Kim, Byeongchang and Lee, Hyunmin and Kim, Gunhee},
  booktitle={Proceedings of the 2019 Conference of the North American Chapter of the Association for Computational Linguistics: Human Language Technologies, Volume 1 (Long and Short Papers)},
  pages={119--132},
  year={2019}
}

@inproceedings{elizalde2023clap,
  title={Clap learning audio concepts from natural language supervision},
  author={Elizalde, Benjamin and Deshmukh, Soham and Al Ismail, Mahmoud and Wang, Huaming},
  booktitle={icassp},
  pages={1--5},
  year={2023},
  organization={IEEE}
}

@article{ho2020denoising,
  title={Denoising diffusion probabilistic models},
  author={Ho, Jonathan and Jain, Ajay and Abbeel, Pieter},
  journal={Advances in neural information processing systems},
  volume={33},
  pages={6840--6851},
  year={2020}
}

@inproceedings{ho2021classifier,
  title={Classifier-Free Diffusion Guidance},
  author={Ho, Jonathan and Salimans, Tim},
  booktitle={NeurIPS 2021 Workshop on Deep Generative Models and Downstream Applications}
}

@article{xu2025uniflow,
  title={UniFlow-Audio: Unified Flow Matching for Audio Generation from Omni-Modalities},
  author={Xu, Xuenan and Mei, Jiahao and Zheng, Zihao and Tao, Ye and Xie, Zeyu and Zhang, Yaoyun and Liu, Haohe and Wu, Yuning and Yan, Ming and Wu, Wen and others},
  journal={arXiv preprint arXiv:2509.24391},
  year={2025}
}

@inproceedings{liu2023audioldm,
  title={AudioLDM: Text-to-Audio Generation with Latent Diffusion Models},
  author={Liu, Haohe and Chen, Zehua and Yuan, Yi and Mei, Xinhao and Liu, Xubo and Mandic, Danilo and Wang, Wenwu and Plumbley, Mark D},
  booktitle={International Conference on Machine Learning},
  pages={21450--21474},
  year={2023},
  organization={PMLR}
}

@inproceedings{peebles2023scalable,
  title={Scalable Diffusion Models with Transformers},
  author={Peebles, William and Xie, Saining},
  booktitle={2023 IEEE/CVF International Conference on Computer Vision (ICCV)},
  pages={4172--4182},
  year={2023},
  organization={IEEE}
}

@inproceedings{salamon2017scaper,
  title={Scaper: A library for soundscape synthesis and augmentation},
  author={Salamon, Justin and MacConnell, Duncan and Cartwright, Mark and Li, Peter and Bello, Juan Pablo},
  booktitle={2017 IEEE Workshop on Applications of Signal Processing to Audio and Acoustics (WASPAA)},
  pages={344--348},
  year={2017},
  organization={IEEE}
}

@inproceedings{songdenoising,
  title={Denoising Diffusion Implicit Models},
  author={Song, Jiaming and Meng, Chenlin and Ermon, Stefano},
  booktitle={International Conference on Learning Representations}
}

\end{document}